\documentclass[final,3p,times]{elsarticle}

\usepackage{amssymb}

\usepackage{threeparttable}
\usepackage{booktabs}
\usepackage{multirow}
\usepackage[noend]{algpseudocode}
\usepackage{algorithmicx,algorithm}

\usepackage{hyperref}
\hypersetup{hidelinks}
\usepackage{amsmath}
\usepackage{subfig}
\usepackage{color} 

\usepackage{verbatim}

\begin{document}

\begin{frontmatter}

\title{Transliterated zero-shot domain adaptation for automatic speech recognition}

\author[label1,label2]{Han Zhu}
\ead{zhuhan@hccl.ioa.ac.cn}
\author[label1]{Gaofeng Cheng}
\ead{chenggaofeng@hccl.ioa.ac.cn}
\author[label1,label2]{Qingwei Zhao}
\ead{zhaoqingwei@hccl.ioa.ac.cn}
\author[label1,label2]{Pengyuan Zhang}
\ead{zhangpengyuan@hccl.ioa.ac.cn}

\affiliation[label1]{organization={Key Laboratory of Speech Acoustics and Content Understanding, Institute of Acoustics, Chinese Academy of Sciences},
            city={Beijing},
            country={China}}
\affiliation[label2]{organization={University of Chinese Academy of Sciences},
            city={Beijing},
            country={China}}

\begin{abstract}
The performance of automatic speech recognition models often degenerates on domains not covered by the training data. Domain adaptation can address this issue, assuming the availability of the target domain data in the target language. However, such assumption does not stand in many real-world applications. To make domain adaptation more applicable, we address the problem of zero-shot domain adaptation (ZSDA), where target domain data is unavailable in the target language. Instead, we transfer the target domain knowledge from another source language where the target domain data is more accessible. To do that, we first perform cross-lingual pre-training (XLPT) to share domain knowledge across languages, then use target language fine-tuning to build the final model. One challenge in this practice is that the pre-trained knowledge can be forgotten during fine-tuning, resulting in sub-optimal adaptation performance. To address this issue, we propose transliterated ZSDA to achieve consistent pre-training and fine-tuning labels, leading to maximum preservation of the pre-trained knowledge. Experimental results show that transliterated ZSDA relatively decreases the word error rate by 9.2\% compared with a wav2vec 2.0 baseline. Moreover, transliterated ZSDA consistently outperforms self-supervised ZSDA and performs on par with supervised ZSDA, proving the superiority of transliteration-based pre-training labels.
\end{abstract}

\begin{keyword}

Automatic speech recognition \sep domain adaptation \sep zero-shot  \sep cross-lingual pre-training
\end{keyword}

\end{frontmatter}


\section{Introduction}
Automatic speech recognition (ASR) aims to automatically transcribe human speech into text, serving as a crucial component in numerous human-machine interaction systems such as chatbots and intelligent assistants. In recent years, significant progress has been made in the field of ASR~\citep{li2022recent}. However, applying ASR models on domains that are not covered by the training set is still challenging due to the distribution shift between training and testing set~\citep{bell2020adaptation}. For example, an ASR model trained on the conversational telephone speech domain would perform unsatisfactorily on the reading speech domain.
Conventional domain adaptation approaches~\citep{manohar2018teacher,sun2017unsupervised,zhu2022boosting,KHEDDAR2023110851,NADERI2023110814,zhang2022data,li2023intelligent} can boost the target domain performance, which assumes the availability of target domain data in the target language. However, such assumption does not always stand due to realistic constraints. For examples, it would be hard to collect target domain data for low-resource languages.

To expand the application scope of domain adaptation, we address the \emph{zero-shot domain adaptation (ZSDA)}~\citep{peng2018zero} problem where we instead leverage target domain data in other languages, thus avoiding the requirement of target domain data in the target language. The comparison between ZSDA and conventional domain adaptation is shown in \autoref{tab:zsda}. We propose to realize ZSDA by \emph{cross-lingual pre-training and target language fine-tuning}. During cross-lingual pre-training (XLPT), since the same set of parameters is used to model source and target languages, the domain knowledge is implicitly shared across languages. Then, we refine the model for the target language by fine-tuning the pre-trained model on only the target language data, which transforms the pre-trained model into the target language ASR model. With this framework, the domain knowledge of the source language can be transferred to the target language.

The effectiveness of the proposed framework for ZSDA largely depends on the selected XLPT method. With self-supervised or supervised XLPT, we can implement self-supervised or supervised ZSDA variants. However, these variants are flawed in terms of performance or cost. We illustrate the details as follows.

\begin{table}[t!]
  \centering
  \caption{Comparison between zero-shot domain adaptation and conventional domain adaptation.}
    \begin{tabular}{lcccccc}
    \toprule
    \multirow{2}[4]{*}{Method} & \multicolumn{2}{c}{Training data} & \multicolumn{2}{c}{Adaptation data} & \multicolumn{2}{c}{Testing data} \\
\cmidrule{2-7}          & Domain & Language & Domain & Language & Domain & Language \\
    \midrule
    w/o domain adaptation & source & target & -     & -     & target & target \\
    Conventional domain adaptation & source  & target & target  & target & target & target \\
    Zero-shot domain adaptation & source  & target & target  & source & target & target \\
    \bottomrule
    \end{tabular}%
  \label{tab:zsda}%
\end{table}%

Self-supervised XLPT~\citep{conneau21_interspeech,babu22_interspeech,zhang2023google} jointly learns self-supervised representations for multiple languages, benefiting low-resource languages by transferring knowledge from other languages.
The corresponding pre-training labels comprise various acoustic proprieties (e.g., fundamental frequency, formants, and amplitude)~\citep{choi2022opening}, making the pre-trained model useful in diversified downstream tasks. However, when we only aim at the ASR task where the fine-tuning labels are usually graphemic transcriptions, the difference between pre-training and fine-tuning labels causes considerable representation change during fine-tuning~\citep{pasad2021layer,zhu22c_interspeech}, leading to severe forgetting of the pre-trained knowledge. Consequently, the adaptation performance would be sub-optimal. 
In terms of supervised XLPT, an obvious disadvantage is the requirement for annotation in the source language, which increases the cost of domain adaptation.

To avoid annotating the source language while maintaining consistency between pre-training and fine-tuning labels, we introduce \emph{transliterated ZSDA}, which utilizes \emph{transliterated XLPT}. In transliterated XLPT, pre-training labels for source and target language are transliterations and transcriptions, respectively. Transliterations are graphemic labels in the writing system of the target language, which is generated by decoding the source language speech with a target language ASR model. Since all pre-training and fine-tuning labels are in the same writing system, the pre-trained knowledge forgetting issue is thus alleviated. Moreover, since we generate transliterations in the pseudo-labeling style, annotation in the source language is no longer required.

One concern in transliterated ZSDA is that the transliteration-based pre-training label can be inaccurate, i.e., it can not represent the content of the speech well. To address this issue, on the one hand, we propose to improve the the quality of transliterations with a curriculum XLPT scheme that pre-trains the model with low-level labels, i.e., self-supervised labels, before pre-training on the high-level labels, i.e., transliterations. On the other hand, we use continuous pseudo-labeling~\citep{higuchi2021momentum} to continuously improve the quality of transliterations during pre-training. Moreover, we adopt the "shared-hidden-layer"~\citep{huang2013cross} architecture, which shares representation extraction layers for both languages while keeping separate classifiers for each language, to tackle the token distribution difference between different transcriptions and transliterations. Finally, to ensure positive knowledge sharing across languages, the source language is selected as a language close to the target language.

Experimental results showed that transliterated ZSDA can realize 9.2\% relative WER (word error rate) reduction on the target domain, consistently outperforming self-supervised ZSDA. Moreover, it achieves competitive performance with the supervised ZSDA even though it does not utilize ground-truth annotation of the source language. We summarize the primary contributions as follows. In this work, we propose:
\begin{itemize}
\item \textbf{Transliterated ZSDA}. We propose to realize ZSDA by sharing domain knowledge across languages via XLPT and refining the target language performance via target language fine-tuning. To preserve the shared domain knowledge after pre-training, we introduce transliterations as pre-training labels and formulate transliterated ZSDA, achieving consistent pre-training and fine-tuning labels, thus minimizing domain knowledge forgetting.

\item \textbf{Techniques to facilitate transliterated ZSDA}. To ensure the effectiveness of knowledge sharing in XLPT, on the one hand, we improve the accuracy of the transliteration-based pre-training labels with curriculum XLPT and continuous pseudo-labeling, on the other hand, we tackle the token distribution differences between transcriptions and transliterations with separated classifiers.

\item \textbf{Back-transliteration based quality metric for transliterations}. To monitor the quality of the pre-training labels, we evaluate the accuracy of transliterations by measuring the similarity between the speech-transliteration pairs with a back-transliteration based approach.

\end{itemize}

The rest of the paper is organized as follows. In \autoref{sec:related}, we review related works. Then, we formulate and analyze the proposed framework for ZSDA in \autoref{sec:problem}. The proposed transliterated ZSDA approach is introduced in \autoref{sec:proposed}. We describe experimental settings in \autoref{sec:experiment}, then present experimental results in \autoref{sec:results}. Finally, \autoref{sec:conclusion} concludes the paper.

\section{Related works}
\label{sec:related}

There are extensive efforts to address the domain adaptation problem for ASR. Typical practices include domain-adversarial training~\citep{ganin2015unsupervised,sun2017unsupervised} and semi-supervised learning based approaches~\citep{hsu21_interspeech,zhu2022boosting,khurana2021unsupervised,higuchi2021momentum}. Domain-adversarial training aims to learn domain-invariant representations using unlabeled target domain data, so that the good performance on the source domain can generalize well to the target domain. Semi-supervised learning based approaches train the model on unlabeled target domain data with self-supervised labels or pseudo transcriptions as supervisions. These approaches require unlabeled target domain data in the target language for adaptation. However, such data can be unavailable in some real-world applications, e.g., ASR for low-resource languages. Therefore, we should consider the new problem: zero-shot domain adaptation (ZSDA), which does not require target domain data in the target language. Note that ZSDA still needs target domain data in the source language.

When the target domain data can be created through simulation, we can achieve ZSDA by generating target domain data and then training on it~\citep{li2017large}. Aiming at a more general solution for ZSDA, ~\citet{abad2020cross} proposes to leverage cross-lingual target domain data. Specifically, on the basis of a multi-lingual ASR model trained on multi-lingual source domain data, the shared hidden layers are fine-tuned with source language target domain data. Afterward, the multi-lingual model can generalize better on the target language target domain data. The disadvantage of this approach is the additional requirement for labeled source language data in both source and target domains. In contrast, transliterated ZSDA only requires unlabeled source language data in the target domain.

One unique aspect of the transliterated ZSDA approach is that pre-training and fine-tuning labels are in the same writing system, which is achieved via transliteration. Transliteration was also used in previous works~\citep{khare2021low,thomas2020transliteration} to improve low-resource ASR. A major difference between our work and these works is that they did not address the domain adaptation problem. Additionally, \citet{khare2021low} relies on a text-based transliteration library to obtain transliterations from transcriptions, which demands source language transcriptions and limits the application scope within the supported languages of such library. \citet{thomas2020transliteration} generates transliterations only once with a target language ASR model trained from random initialization, limiting the accuracy of the transliteration and the final performance.

\section{Address ZSDA problem with cross-lingual pre-training and target language fine-tuning}
\label{sec:problem}

\subsection{Formulation}
\label{sec:problem_formulation}
\begin{figure}[!t]
  \centering%
  \includegraphics[width=0.7\columnwidth]{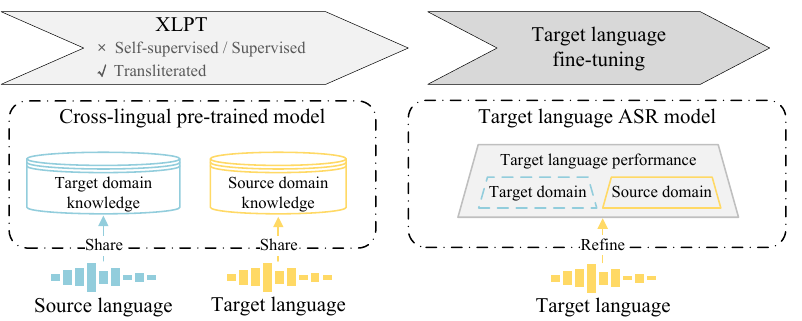}
\caption{The diagram of the proposed ZSDA framework: cross-lingual pre-training (XLPT) and target language fine-tuning. The proposed transliterated ZSDA method is a special case of this framework, where we use transliterated XLPT.}
\label{fig:illustration_framework}
\end{figure}

The task of ASR can be formulated as $\mathbf{y} = f_{\theta}(\mathbf{x})$.
Suppose we have a labeled source domain dataset $\mathbb{L} = \{(\mathbf{x}_0, \mathbf{y}_0), \ldots\, (\mathbf{x}_M, \mathbf{y}_M)\}$ in the target language, and an unlabeled target domain dataset $\mathbb{U} =\{\mathbf{x}_0, \ldots, \mathbf{x}_N\}$ in the source language. The goal of ZSDA is: given available datasets $\mathbb{U} \cup \mathbb{L}$, optimizing the model $\theta$ so that the prediction accuracy of $y$ on the target domain can be improved. Our framework to address the ZSDA problem is XLPT on $\mathbb{U} \cup \mathbb{L}$ followed by target language fine-tuning on $\mathbb{L}$, which is illustrated in \autoref{fig:illustration_framework}. For example, say the target language dataset $\mathbb{L}$ is in the domain of conversational telephone speech, while the source language dataset $\mathbb{U}$ is in the domain of reading speech. With the above framework, the reading speech domain knowledge can be transferred to the target language, thus improving the performance of the target language on the reading speech domain. 

\subsection{Analysis}

\label{sec:problem_analysis}

In the proposed framework, the final performance relies on the effectiveness of domain knowledge transfer across languages. In the following, we analyze the factors that can affect the effectiveness of domain knowledge transfer. 

During pre-training, desired knowledge transfer relies on effective knowlege sharing among languages. Firstly, from the perspective of pre-training data, the \emph{similarity between the source and target languages} is critical. If the source language is not sufficiently similar to the target language, transferring from such source language may even hinder the performance of the target language, a phenomenon known as negative transfer~\citep{wang2019characterizing}. Secondly, regarding the pre-training label, the \emph{accuracy of pre-training labels} in the unlabeled source language should be guaranteed for the effectiveness of pre-training. Thirdly, concerning the model architecture, \emph{proper parameter sharing among languages} is important. For example, in self-supervised XLPT where pre-training labels for all languages are unified, we usually share all parameters across languages. In contrast, in supervised XLPT where pre-training labels for different languages are in different writing systems, the typical practice is to share the representation extraction layers while keeping separate final linear layers with the assumption that the representation extraction function can be shared across languages~\citep{huang2013cross}. 

During fine-tuning, the effectiveness of knowledge transfer is negatively related to the extent of representation change, as the severe representation change leads to critical distortion of the pre-trained knowledge and leaves the target domain performance at risk (detailed analysis is performed in \autoref{sec:knowlege_preservation_analysis}). For example, in self-supervised XLPT, the representation differences between the pre-trained and fine-tuned model lead to unsatisfactory domain knowledge preservation. Some works utilize phoneme labels during self-supervised XLPT to guide the pre-trained model to be ASR-specific~\citep{wang2021unispeech,zhang2022cross}, thus decreasing the representation change during fine-tuning. Nonetheless, there are still considerable differences between the phonemic pre-training and graphemic fine-tuning labels. To this end, we identify that ideal knowledge preservation requires \emph{graphemic pre-training labels} in the same writing system with the fine-tuning data. Moreover, given the assumption that the source language data is unlabeled, the generation of such pre-training labels on the source language data should be unsupervised, i.e., not rely on the ground-truth transcription.

\section{Proposed approach: Transliterated ZSDA}
\label{sec:proposed}

Following the analysis in \autoref{sec:problem_analysis}, we propose transliterated ZSDA under the proposed framework for ZSDA to boost the domain knowledge transfer. The overall procedure of transliterated ZSDA is shown in Algorithm~\ref{alg:T-ZSDA}. We briefly describe the primary designs as follows.

\begin{figure}
\centering
\begin{minipage}[t]{.65\linewidth}
\begin{algorithm}[H]
	\caption{Transliterated ZSDA algorithm.}
	\begin{small}
	
	{\bf Input:}
	Labeled target language dataset in the source domain $\mathbb{L}$ \\
	{\bf Output:} 
	Target language ASR model $\mathcal{M}_\text{ASR}$.
	\begin{algorithmic}[1]
	
		\State Find a source language close to target language.
            \State Prepare unlabeled source language dataset in the target domain $\mathbb{U}$;
            \State Randomly initialize or reuse an existing self-supervised model $\mathcal{M}_\text{SSL}$;
		\Repeat 
            \Comment{\textit{\textbf{Begin curriculum XLPT}}}
		\State Draw a batch $\mathbf{B}$ from $\mathbb{L} \cup \mathbb{U}$;
            \State Compute self-supervised loss $\mathcal{L}_{\text{SSL}}$ on $\mathbf{B}$ to update the model $\mathcal{M}_\text{SSL}$;
		\Until{maximum updates for curriculum XLPT are reached} 
            \Comment{\textit{\textbf{End}}}
  		\State Add random initialized linear layer on $\mathcal{M}_\text{SSL}$ to produce $\mathcal{M}_\text{ASR}$;
            \Repeat 
            \Comment{\textit{\textbf{Begin transliterated XLPT}}}
            \State Draw a batch $\mathbf{B}_{L}$ from $\mathbb{L}$;
            \State Compute supervised loss $\mathcal{L}_{\text{sup}}$ on $\mathbf{B}_{L}$ to update the model $\mathcal{M}_\text{ASR}$ ;
            \Until{maximum updates of the seeding stage are reached}
            \Repeat
            \State Draw batches $\mathbf{B}_{L}$, $\mathbf{B}_U$ of the same size from $\mathbb{L}$ and $\mathbb{U}$;
            \State Compute supervised loss $\mathcal{L}_{\text{sup}}$ on $\mathbf{B}_{L}$, and pseudo-labeling loss $\mathcal{L}_{\text{PL}}$ on $\mathbf{B}_{U}$;
            \State Update the model $\mathcal{M}_\text{ASR}$ with the combination of both losses $\mathcal{L}_{\text{sup}} + \mathcal{L}_{\text{PL}}$;
            \Until{maximum updates of the pseudo-labeling stage are reached}
            \Comment{\textit{\textbf{End}}}
            \State Randomly initialize the final linear layer of $\mathcal{M}_\text{ASR}$
            \Repeat
            \Comment{\textit{\textbf{Begin target language fine-tuning}}}
            \State Draw a batch $\mathbf{B}_{L}$ from $\mathbb{L}$;
            \State Compute supervised loss $\mathcal{L}_{\text{sup}}$ on $\mathbf{B}_{L}$ to update the model $\mathcal{M}_\text{ASR}$;
            \Until{maximum updates for target language fine-tuning are reached}
            \Comment{\textit{\textbf{End}}}
	\end{algorithmic}
	\end{small}
	\label{alg:T-ZSDA}
\end{algorithm}
\end{minipage}
\end{figure}

To maximize domain knowledge preservation after fine-tuning, transliterated ZSDA uses transliterations as pre-training labels for the source language so that all pre-training and fine-tuning labels are consistent. The generation and utilization of the transliteration adopt the pseudo-labeling method, thus enabling the unsupervised generation of pre-training labels for the unlabeled source language data.

The following designs ensure effective domain knowledge sharing during pre-training.
Firstly, transliterated ZSDA utilizes a source language closely related to the target language in XLPT. Specifically, the source and target languages are in the same language family.
Secondly, we propose a curriculum XLPT scheme that performs XLPT with self-supervised pre-training labels before the transliteration-based pre-training labels, thus improving the accuracy of the generated transliterations.
Thirdly, we propose adopting the "shared-hidden-layer" architecture to tackle the token distribution mismatch between transcriptions and transliterations, realizing proper parameter sharing across languages.

\subsection{XLPT method of transliterated ZSDA: transliterated XLPT}
\label{sec:T-XLPT}

As discussed in \autoref{sec:problem_analysis}, the ideal pre-training labels should be in the same writing system with the fine-tuning labels. Fortunately, transliterations, which represent the content of one language with the writing system of another language, can satisfy this requirement. Therefore, we use transliterations as pre-training labels, formulating transliterated XLPT.

Conventionally, transliterations are generated by converting text from a language's writing system to another language's writing system~\citep{deselaers2009deep}, i.e., the operation is only conducted on the textual level. Considering only the unlabeled speech is available in the source language, we instead generate transliterations similarly with pseudo-labeling~\citep{kahn2020self,higuchi2021momentum,WEN2022109589,ZHANG2021107340}.

\begin{figure}[t!]
  \centering%
  \includegraphics[width=0.45\columnwidth]{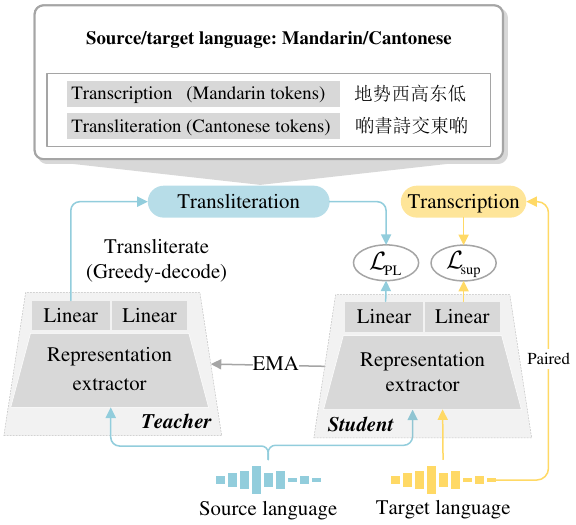}
\caption{Illustration transliterated XLPT. In this example, the target language is Cantonese, whereas the source language is Mandarin. To illustrate how the transliteration reflects the content of the source language speech, we provide the original speech of the source language and the speech synthesized with the transliteration in \url{https://zhu-han.github.io/transliteration}.}
\label{fig:T-XLPT}
\end{figure}

Before pseudo-labeling, a seed ASR model should be prepared with supervised learning, which is used to generate transliterations in the following pseudo-labeling stage. Specifically, we train a supervised seed model on the labeled target language data set $\mathbb{L}$ with the supervised loss:
\begin{equation}
\label{equ:supervised}
\mathcal{L}_{\text{sup}} = \operatorname{CTC}(\mathbf{y}, f_{\mathbf{\theta}_{t}}(a(\mathbf{x})),  (\mathbf{x}, \mathbf{y}) \in \mathbb{L}
\end{equation}
where $\theta_{t}$ is the model in the $t$-th training update, $f$ denotes the forward process of the model, $a(\cdot)$ is the data augmentation function. The data augmentation strategy for follows \citet{baevski2020wav2vec}, i.e., masking in both time and channel dimensions, similar to SpecAugment~\citep{park2019specaugment}. $\operatorname{CTC}(\cdot)$ denotes the connectionist temporal classification (CTC)~\citep{graves2006connectionist} loss, which can be formulated as the negative log probability:
\begin{equation}
\label{equ:ctc}
    \operatorname{CTC}(\mathbf{y}, \mathbf{z}) =
    - \log P(\mathbf{y}|\mathbf{z})
\end{equation}
where $\mathbf{z}$ is the output of the ASR model.

After the preparation of the supervised seed model, the second stage is the continuous pseudo-labeling stage (shown in \autoref{fig:T-XLPT}), which shares domain knowledge across languages and gradually improves the quality of transliterations. Continuous pseudo-labeling~\citep{higuchi2021momentum} is a kind of pseudo-labeling method that update the pseudo-labels continuously during training. Resuming from the supervised seed model, the student model $\theta_{t}$ is then trained with the above supervised loss and an additional pseudo-labeling loss. The additional pseudo-labeling loss in the $t$-th round of update is computed on the unlabeled source language dataset $\mathbb{U}$ as:
\begin{equation}
\label{equ:unlabel}
\mathcal{L}_{\text{PL}} = \operatorname{CTC}(\hat{\mathbf{y}}, f_{\mathbf{\theta}_{t}}(a(\mathbf{x})), \mathbf{x} \in \mathbb{U}
\end{equation}
where $\hat{\mathbf{y}}$ denotes the transliteration. Language model should not be involved in generating transliterations, as the transliteration only relies on the pronunciation and is semantically meaningless. Therefore, we use greedy-decoding to generate the transliteration:
\begin{equation}
    \label{equ:transliterate}
    \hat{\mathbf{y}}=\operatorname{greedy-decode} f_{\xi_{t}}(\mathbf{x}), \mathbf{x} \in \mathbb{U}
\end{equation}
which involves two models: The teacher model $\xi_{t}$ is the model to generate transliterations and the student model $\theta_{t}$ is the model to utilize the transliterations. The teacher model is updated as the exponential moving average (EMA) of the student model for continuous improvement of transliterations:
\begin{equation}
 \xi_{t} = \alpha \xi_{t-1} + (1-\alpha) \theta_{t}  
\end{equation}
where $\alpha \in (0,1)$ is the decay factor. 

The greedy-decoding of CTC-based ASR model contains three operations: find the token with maximum probability in each position, merging consecutive identical tokens, and removing blank tokens.

As shown in \autoref{fig:T-XLPT}, when the target and source languages are Cantonese and Mandarin, respectively, the transliteration of the Mandarin speech is in the writing system of Cantonese. Although these transliterations are semantically meaningless, they can reflect the content of the Mandarin language speech since they are generated according to the pronunciation of the Mandarin speech.
Since the transliteration is written with the target language tokens, all pre-training and fine-tuning labels are in the writing system of the target language. Such consistency of labels leads to less representation change and more pre-trained knowledge preservation during fine-tuning.

Since the transliteration of the source language can have quite a different label distribution from the transcription of the target language, sharing the label classifier, i.e., the final linear layer of the ASR model, would be sub-optimal. Therefore, we divide the classifier into two branches, each accounting for one language. The parameters of both branches are initialized from the supervised seed model. Such design follows the assumption that the representation extraction can be shared among languages~\citep{huang2013cross}, while the label classifier may not.

\subsection{Boosting the quality of transliterations: curriculum XLPT}

Since we generate transliterations in a pseudo-labeling fashion, the accuracy of transliterations can not be guaranteed. By accuracy, we mean how well the label can reflect the content of the speech. Inaccurate transliterations can degenerate the effectiveness of knowledge sharing. Unfortunately, the supervised seed model suffers from the distribution shift when generating transliterations for the source language speech, as it is only trained on target language data. Thus, the generated transliterations would be of low quality.

To boost the accuracy of transliteration, we propose the curriculum XLPT scheme that initializes the model with self-supervised XLPT, i.e., self-supervised pre-training on both source and target language data. Since self-supervised pre-training on out-of-domain data can improve the performance on the out-of-domain distribution~\citep{hsu21_interspeech}, when using the pre-trained model as the initialization for transliterated XLPT, the quality degradation issue caused by distribution shift can be effectively alleviated.

\subsection{Quality metric of transliteration: back-transliterated CTC Loss}

In the above, we assume the curriculum XLPT scheme and the continuous pseudo-labeling stage in transliterated XLPT can improve the quality of transliterations. Naturally, we want to compare the quality of transliterations with and without these two techniques. However, different from the transcription-based pseudo-labels that can be directly measured with word error rate (WER), evaluating the quality of the generated transliterations would be hard.

\begin{figure}[t!]
  \centering%
  \includegraphics[width=0.7\columnwidth]{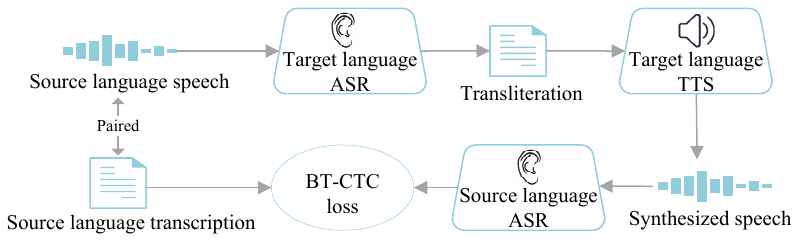}
\caption{Computation procedure of BT-CTC loss. BT-CTC loss can measure the similarity between transliteration and source language speech.}
\label{fig:BT-CTC}
\end{figure}

To fill this gap, we propose a quality metric called back-transliterated CTC loss (BT-CTC) to measure the similarity between the transliterations and the original speech, whose computation procedure is similar to the back-transliteration~\citep{knight1997machine} operation in machine translation.
We illustrate the procedure in \autoref{fig:BT-CTC}, and describe the details as follows.

Given a source language speech $\mathbf{x}$, we generate its transliteration $\hat{\mathbf{y}}$ in the writing system of the target language with the transliterated XLPT model (like a target language ASR model) as in \autoref{equ:transliterate}. Then, we synthesize speech of the transliteration with a target language text-to-speech (TTS) model:
\begin{equation}
    \hat{\mathbf{x}}=\operatorname{TTS}_\text{target}(\hat{\mathbf{y}})
\end{equation}

Although the speech $\hat{\mathbf{x}}$ is synthesized with the target language TTS model, it is semantically meaningless to a target language speaker. Instead, it sounds similar to the source language speech $\mathbf{x}$. Recall that accuracy means how well the transliterations can reflect the content of the source language speech. Therefore, the content similarity between the synthesized speech $\hat{\mathbf{x}}$ and the source language speech $\mathbf{x}$ reflect the accuracy of the transliteration.

Since the transcription $\mathbf{y}$ can represent the content of the $\mathbf{x}$, and a source language ASR model can extract the content of the synthesized speech $\hat{\mathbf{x}}$, we use the following CTC loss to measure the similarity between $\hat{\mathbf{x}}$ and $\mathbf{x}$:
 \begin{equation}
\label{equ:back}
\mathcal{L}_{\text{BT-CTC}}\left(\theta\right) = \operatorname{CTC}(\mathbf{y}, f_{\mathbf{\phi}}(\hat{\mathbf{x}}))
\end{equation}
where $\mathbf{\phi}$ is an ASR model of the source language.

\section{Experimental setup}
\label{sec:experiment}

\subsection{Corpus}
We treat Cantonese and Czech as target languages, whereas Mandarin, Russian, and English as source languages. Cantonese and Mandarin are both Sinitic languages within the Sino-Tibetan language family. Czech and Russian are both Balto-Slavic languages within the Indo-European language family. Therefore, we use Mandarin and Russian as source languages for Cantonese and Czech, respectively. English is used as an auxiliary language in ablation study, which is a Germanic language of the Indo-European language family.

To evaluate the ZSDA, the training and testing set of the target language should be in different domains, while the training set of the source language should be in the same domain with the testing set of the target language. Under such constraint, we construct the experimental corpora in \autoref{tab:corpus}, and explain the details as follows.

\begin{table}[htbp]
  \small
  \centering
  \caption{Statistics for corpora of target and source languages.}
    \begin{tabular}{lcccccccccc}
    \toprule
    Language type & \multicolumn{6}{c}{Target language} & \multicolumn{4}{c}{Source language} \\
    \cmidrule(lr){1-1} \cmidrule(lr){2-7} \cmidrule(lr){8-11}
    Language & \multicolumn{3}{c}{Cantonese} & \multicolumn{3}{c}{Czech} & \multicolumn{2}{c}{Mandarin} & Russian & English \\
     \cmidrule(lr){1-1} \cmidrule(lr){2-4} \cmidrule(lr){5-7} \cmidrule(lr){8-9} \cmidrule(lr){10-10} \cmidrule(lr){11-11}
    Split & Train & Dev   & Test  & Train & Dev   & Test  & \multicolumn{2}{c}{Train} & Train & Train \\
     \cmidrule(lr){1-1} \cmidrule(lr){2-2} \cmidrule(lr){3-3} \cmidrule(lr){4-4} \cmidrule(lr){5-5} \cmidrule(lr){6-6} \cmidrule(lr){7-7} \cmidrule(lr){8-9} \cmidrule(lr){10-10} \cmidrule(lr){11-11}
    Corpus & MDCC  & CV    & CV    & CTS   & CV    & CV    & CV    & Aishell-1 & CV    & CV \\
    \midrule
    Duration (hours) & 57.5  & 7.5   & 8.2   & 15.4  & 11.2  & 11.3  & 143.2 & 150.9 & 143.5 & 143.7 \\
    \bottomrule
    \end{tabular}%
  \label{tab:corpus}%
\end{table}%

For Cantonese, the training set is from the MDCC~\citep{yu2022automatic} dataset, while the development and testing sets are from the Common Voice (CV)~\citep{ardila2020common} Cantonese dataset. Both MDCC and CV datasets are reading speech. But the speech of MDCC is from Cantonese audiobooks where the speakers and the recording devices are professional, while non-professional volunteers record the speech of CV with their own recording devices. Therefore, MDCC and CV are in mildly different domains. As for Czech, the training set is from a conversational telephone speech (CTS) Czech corpus \citep{korvas2014free}, while the development and testing datasets are from the Common Voice (CV) Czech corpus. Therefore, the domains of Czech training and testing sets are significantly different, i.e., conversational telephone speech domain and reading speech domain.

Since all testing datasets are constructed from the corresponding language's CV dataset, we also use the CV dataset to construct the unlabeled training sets for source languages. We additionally prepare AISHELL-1~\citep{bu2017aishell}, a reading mandarin corpus, for the ablation study.

All audios are re-sampled to 16kHz and transcripts are pre-processed to upper-case letters and no punctuation except apostrophes. Therefore, all corpora have the same speech and transcription formats.

\subsection{Implementation}

All experiments are conducted with the FAIRSEQ \citep{ott2019fairseq} toolkit. We adopt the model architecture of wav2vec 2.0~\citep{baevski2020wav2vec} in all experiments.

The XLPT procedures in different ZSDA methods are described as follows. For transliterated ZSDA, the training updates of curriculum XLPT and transliterated XLPT are 10k and 30k, respectively. In transliterated XLPT, the first 10k updates are for the seeding stage, and the last 20k is for the pseudo-labeling stage. For a fair comparison, other ZSDA methods also have the same total pre-training updates, i.e., 40k. All XLPT methods adopt a fixed learning rate of $3*10^{-5}$ with the Adam~\citep{kingma2014adam} optimizer. And the effective batch size is 25.6m audio samples.

As for target language fine-tuning in all ZSDA methods, we randomly initialize the final linear layer and inherit other layers from the pre-trained model. Regarding transliterated ZSDA, re-initializing the final linear layer is just for a fair comparison, which does not make a difference in results according to our experiments. The loss function in fine-tuning is the CTC loss. The total fine-tuning updates are 40k. In the first 5k updates, only the final linear layer is trained. The learning rate, optimizer, and batch size are the same as in pre-training.

In terms of evaluation. We use the greedy-decoding method to decode the final ASR model. And the metric to evaluate the performance is word error rate (WER) and character error rate (CER) for Czech and Cantonese, respectively. For simplicity, we use WER for both terms. The range of WER is $(0, 100)$ and lower value denotes better performance.

\section{Experimental results}
\label{sec:results}
\subsection{Comparison of different methods}
\label{sec:comparison}

\begin{table}[htbp]
  \small
  \centering
  \caption{WER of ASR models w/ supervised training or different ZSDA methods (Best results are in bold).}
      \begin{threeparttable}
    \begin{tabular}{lccccc}
    \toprule
    \multirow{2}[4]{*}{Method}   & \multicolumn{2}{c}{Cantonese} & \multicolumn{2}{c}{Czech} & \multirow{2}[4]{*}{Avg} \\
\cmidrule{2-5}          & Dev   & Test  & Dev   & Test  &  \\
    \midrule
    \multicolumn{6}{l}{\textit{Supervised training w/o target domain data}} \\
    Training from scratch & 53.1  & 57.3  & 103.2 & 102.7 & 79.1 \\
    Training from Wav2vec 2.0 initialization & 32.6  & 34.6  & 55.0  & 55.3  & 44.4 \\
    \midrule
    \multicolumn{6}{l}{\textit{ZSDA with unlabeled target domain data}} \\
    Domain-adversarial training & 32.8  & 35.4  & 55.4  & 55.7  & 44.8 \\
    Self-supervised ZSDA & 31.7  & 34.4  & 50.1  & 49.5  & 41.4 \\
    Transliterated ZSDA   & 30.5  & 32.6  & \textbf{48.9} & \textbf{48.7} & \textbf{40.2} \\
    \midrule
    \multicolumn{6}{l}{\textit{ZSDA with labeled target domain data}} \\
    Supervised ZSDA & 30.5  & 32.5  & 52.9  & 52.2  & 42.0 \\
    + curriculum & \textbf{30.2} & \textbf{32.4} & 49.6  & 49.0  & \textbf{40.3} \\
    \bottomrule
    \end{tabular}%
        \begin{tablenotes} 
        \footnotesize   
        \item[$*$] Results in each row are trained with the same number of updates.
    \end{tablenotes}  
        \end{threeparttable}
  \label{tab:main}%
\end{table}%

We report performances of some baseline methods and ZSDA methods in \autoref{tab:main}.
Firstly, we train baseline ASR models without domain adaptation techniques. A straightforward baseline is the CTC-based ASR model trained from random initialization. Furthermore, we utilize the wav2vec 2.0 model pre-trained on LibriSpeech 960h~\citep{panayotov2015librispeech} as the initialization, which is clearly beneficial as shown in \autoref{tab:main}.

Then, we evaluate approaches under the concerned scenario: ZSDA with unlabeled source language data. We utilize the wav2vec 2.0 model as the initialization of all models to speed up convergence and boost performance. A widely adopted approach in conventional domain adaptation is the domain-adversarial training~\citep{sun2017unsupervised} that aims to obtain domain-invariant representations within one language. We apply this approach to see if it can also be used to learn domain-invariant representations across languages. We sweep the weight for domain classification loss and determine the best model according to the development set. We can observe that domain-adversarial training fails to improve performance. Previous work~\citep{liu2019transferable} also demonstrated that domain-adversarial training fails to improve performance when large domain discrepancy exists. Then, under the proposed framework, we utilize self-supervised XLPT to formulate self-supervised ZSDA, which significantly decreases the WER. Furthermore, we switch the pre-training method to transliterated XLPT, formulating transliterated ZSDA and obtaining consistent performance improvement in two target languages, proving the superiority of using transliterations as pre-training labels. As shown in \autoref{tab:error_type}, compared with the model w/o ZSDA (wav2vec 2.0) and the model w/ self-supervised ZSDA, the WER reduction of transliterated ZSDA primarily comes from the decrease of substitution errors, which means the ASR model can transcribe more accurately.

\begin{table}[htbp]
  \centering
  \caption{Detailed error types of ASR models.}
    \begin{tabular}{lcccc}
    \toprule
    Method & WER   & Substitution & Deletion & Insertion \\
    \midrule
    Wav2vec 2.0 & 32.7  & 31.1  & 1.0   & 0.7 \\
    Self-supervised ZSDA & 31.7  & 30.3  & 0.8   & 0.6 \\
    Transliterated ZSDA & \textbf{30.5} & \textbf{29.2} & \textbf{0.8} & \textbf{0.6} \\
    \bottomrule
    \end{tabular}%
  \label{tab:error_type}%
\end{table}%

Finally, we explore the ZSDA with labeled source language data. Wav2vec 2.0 model is also used as the initialization. We utilize supervised XLPT for domain knowledge sharing, formulating supervised ZSDA. Compared with self-supervised ZSDA, supervised ZSDA performs better in the Cantonese language while worse in the Czech language. When we further apply the curriculum XLPT scheme, i.e., use self-supervised XLPT in the first 10k training updates, the performance is consistently improved in both languages. The improved supervised ZSDA is consistently better than self-supervised ZSDA and comparable with transliterated ZSDA. Note that the effectiveness of curriculum XLPT in the supervised setting is expected since the initialization of self-supervised pre-training can improve the downstream performance even when the pre-training is performed on the same data with fine-tuning~\citep{baevski2020wav2vec,hsu2021hubert}. 

To this end, we can conclude that transliterated ZSDA is the most appealing approach since it performs best and does not require transcriptions for the source language data.

\subsection{Importance of consistent pre-training and fine-tuning labels}
\label{sec:consistent}

\begin{figure}[!ht]
  \centering%
  \includegraphics[width=0.5\columnwidth]{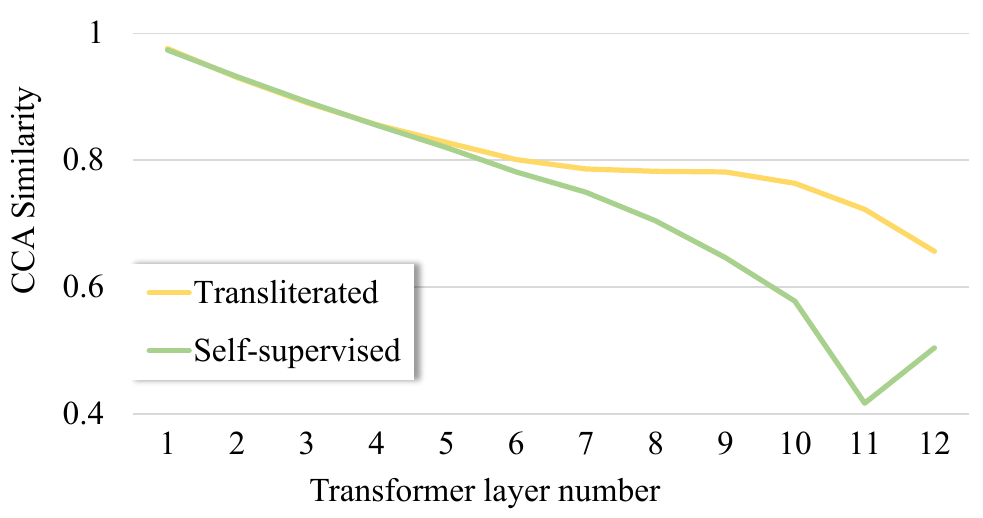}
\caption{Illustration of representation similarity between the pre-trained and fine-tuned models with transliterated or self-supervised XLPT methods. Higher similarity is better.}
\label{fig:CCA}
\end{figure}

To intuitively understand the effect of consistent pre-training and fine-tuning labels, we analyze the representation change during fine-tuning by computing the CCA similarity~\citep{pasad2021layer} between representations of a pre-trained model and the corresponding fine-tuned model in \autoref{fig:CCA}. The lower CCA similarity means the representation changes more significantly during fine-tuning. We can observe that the transliterated XLPT model changes less than the self-supervised XLPT model during fine-tuning, especially for the last few layers.

\begin{figure}[!ht]
    \centering
	\subfloat[Before fine-tuning (self-supervised)] 
	{ \label{fig:SSL_pretrain}
		\includegraphics[height=0.25\columnwidth]{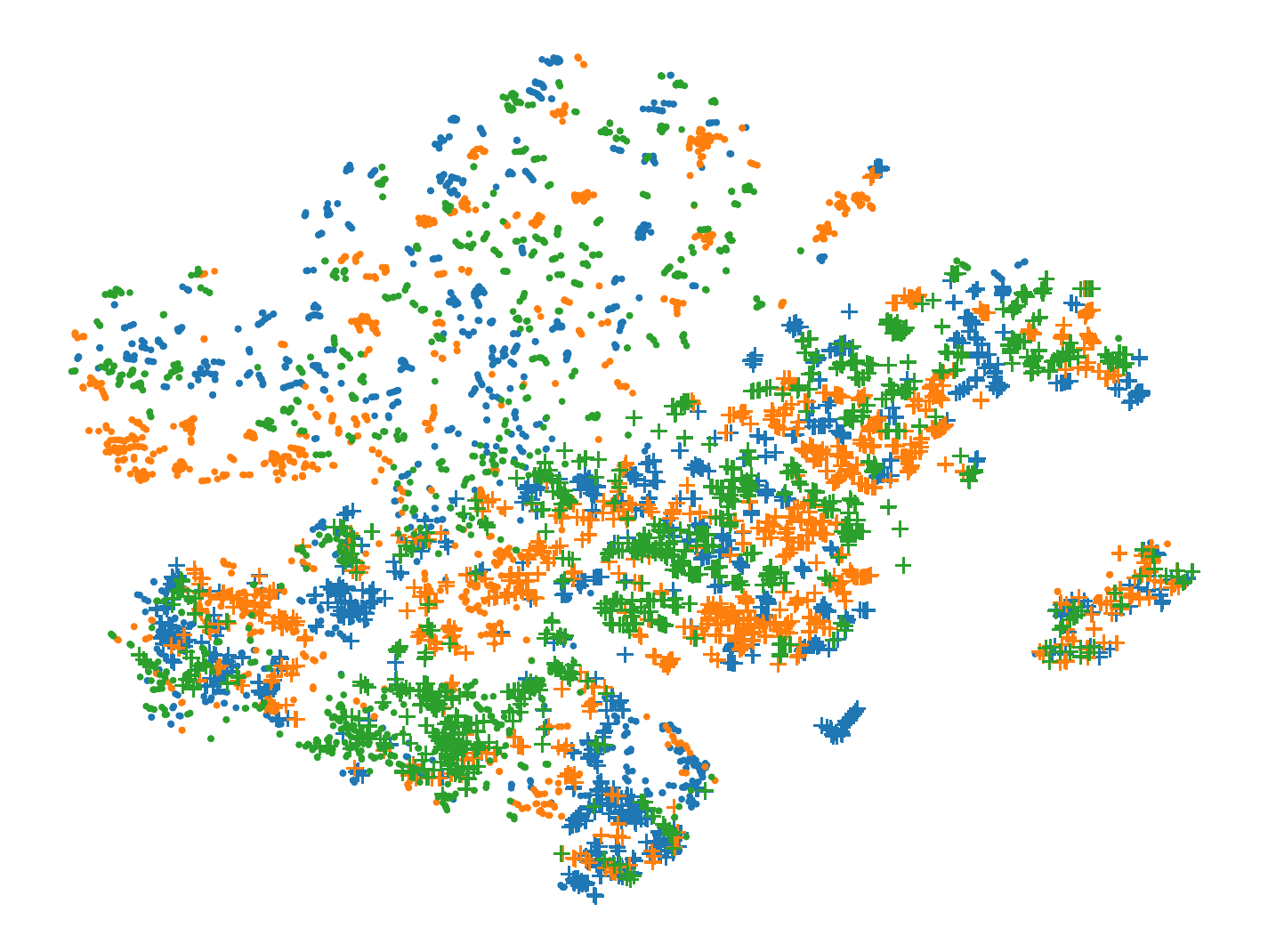}
	}\qquad
	\subfloat[After fine-tuning (self-supervised)] 
	{ \label{fig:SSL_finetune}
		\includegraphics[height=0.25\columnwidth]{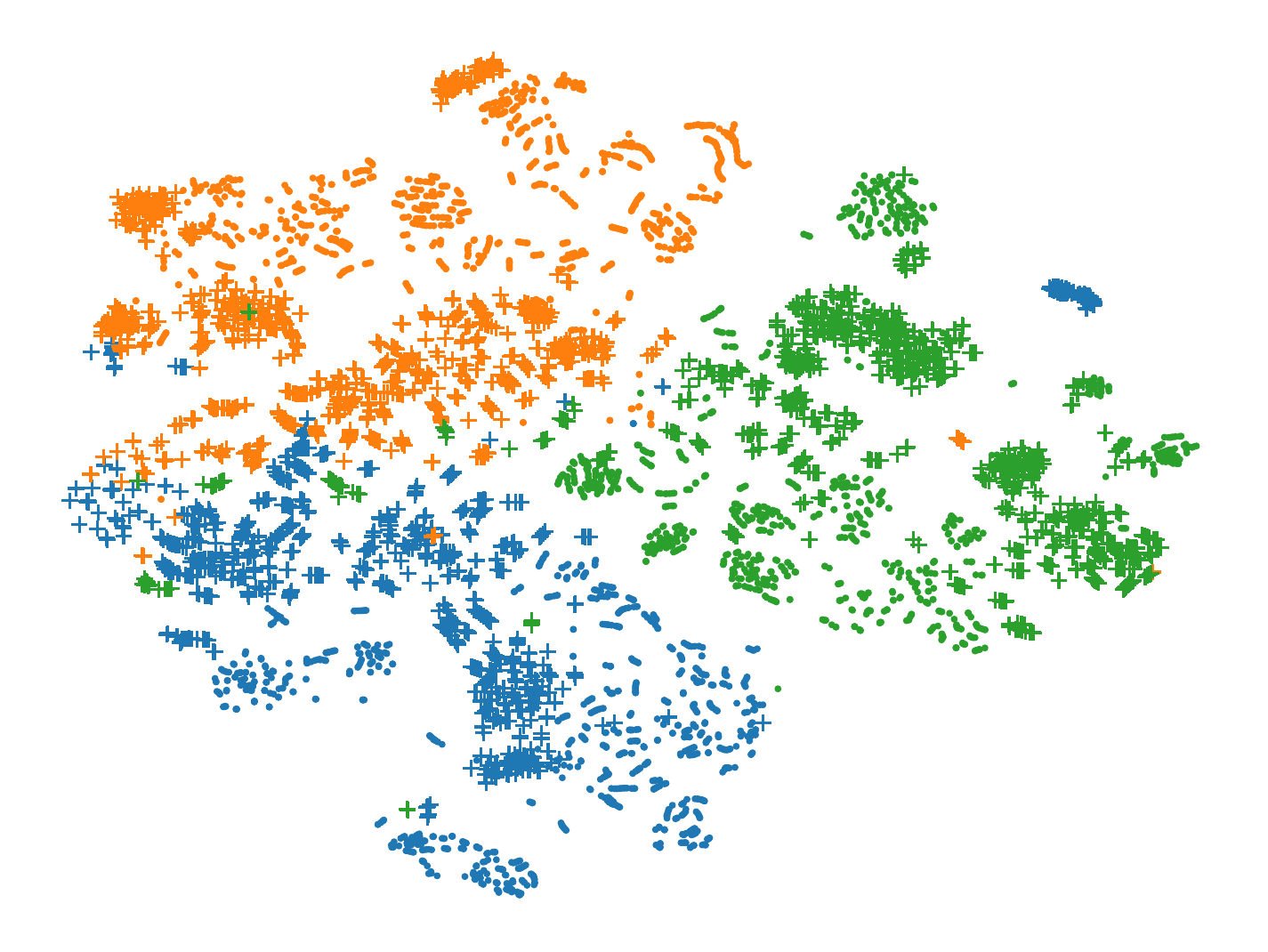}
	}
 
	\subfloat[Before fine-tuning (transliterated)] 
	{ \label{fig:T-XLPT_pretrain}
		\includegraphics[height=0.25\columnwidth]{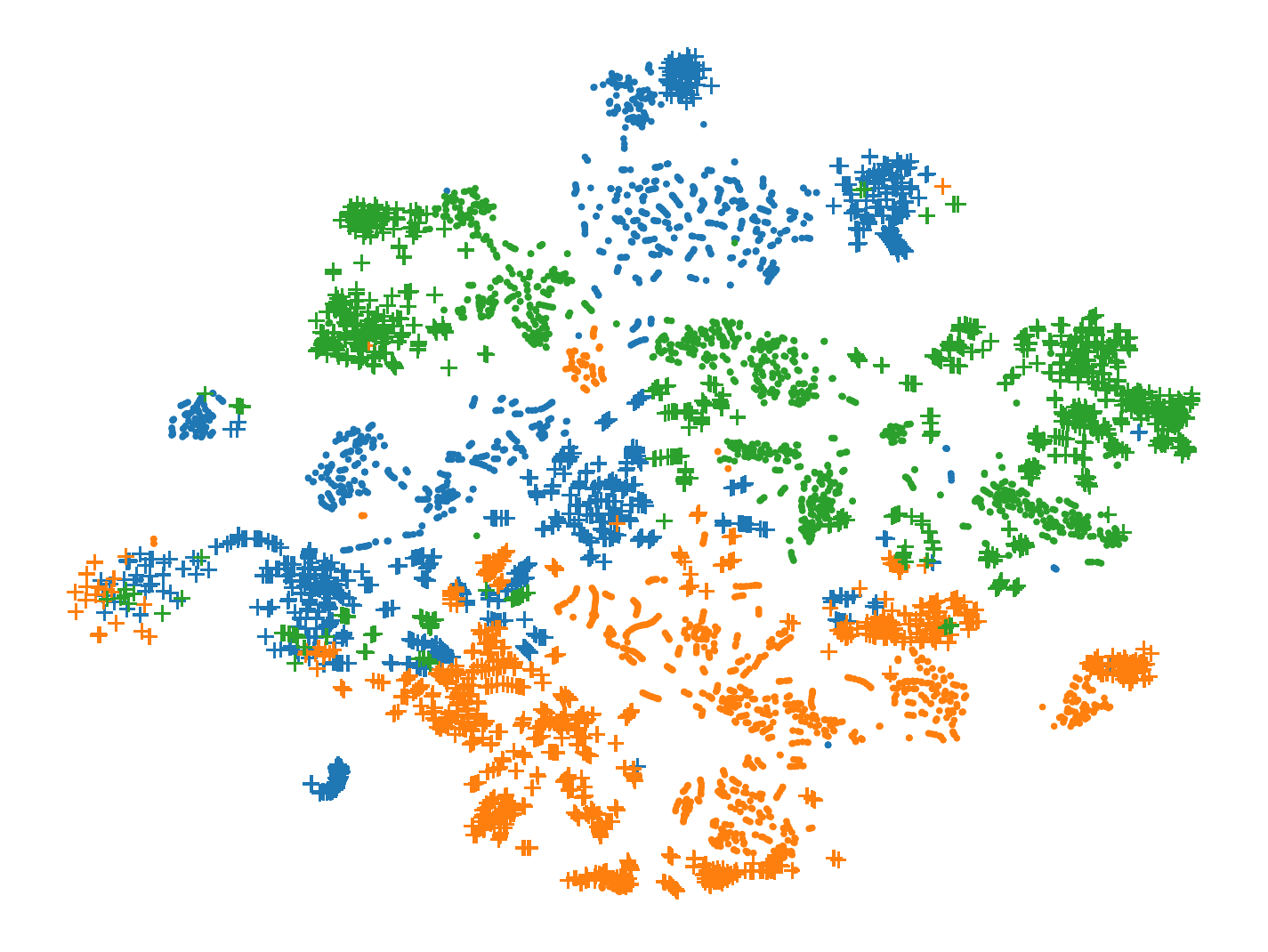}
	}
	\subfloat[After fine-tuning (transliterated)] 
	{ \label{fig:T-XLPT_finetune}
		\includegraphics[height=0.25\columnwidth]{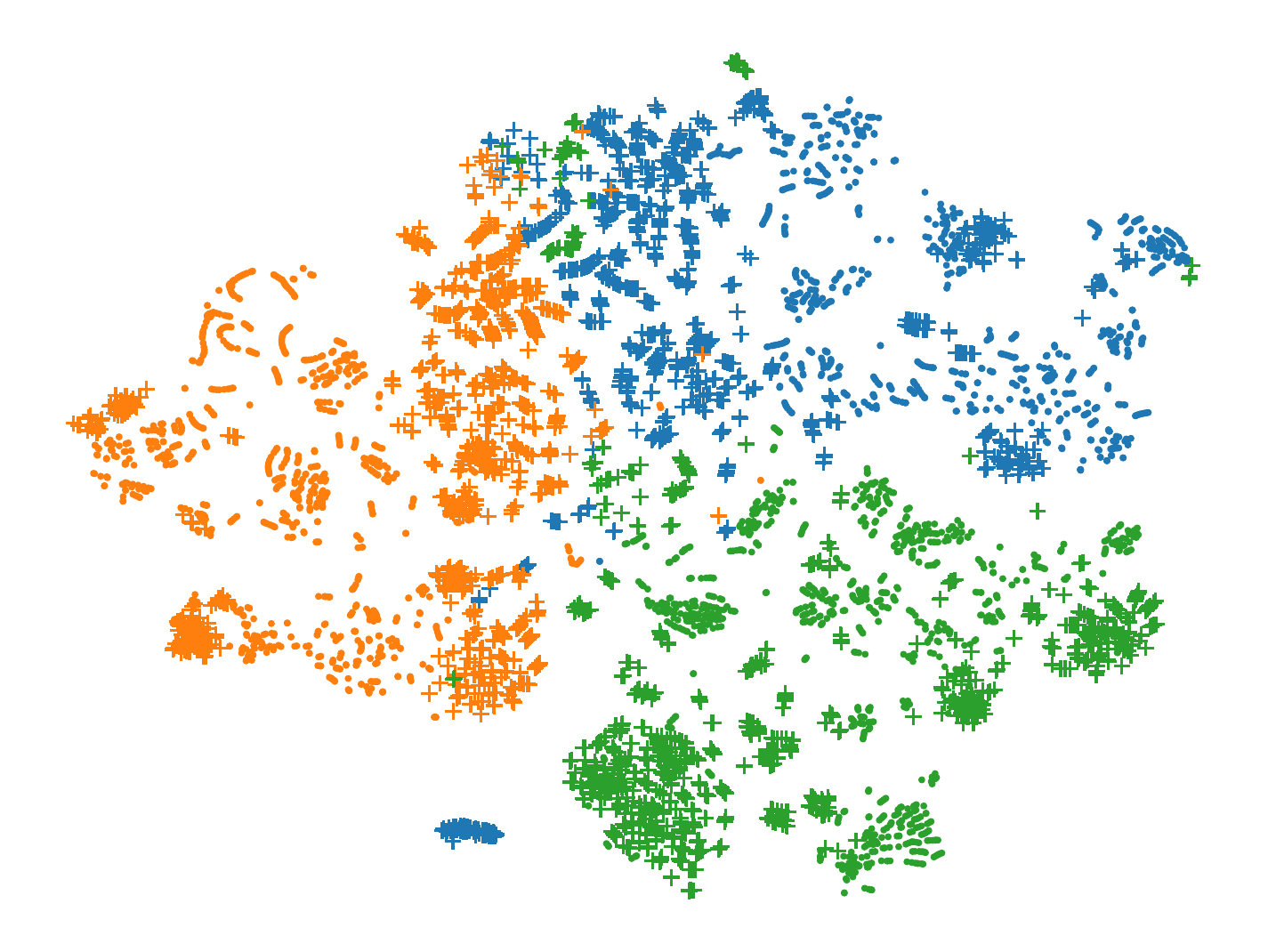}
	}
	\caption{Comparison of representation before and after fine-tuning for self-supervised and transliterated XLPT models. The same color denotes representations aligned to the same output token. The dot symbol $\cdot$ and the plus symbol $+$ denote representations of target and source domain, respectively. \emph{Best viewed in color and zoom in.}} 
	\label{fig:comparison_representation}
\end{figure}

Moreover, we illustrate how the representation is changed during fine-tuning using t-SNE~\citep{vandermaaten08a} visualization in \autoref{fig:comparison_representation}. For the self-supervised XLPT model, representations of different output tokens are not well separated, and representations are not well aligned between source and target domains. The representation pattern changed a lot after fine-tuning: the representations of different tokens are clearly separated with good alignment between the source and target domain. In contrast, representations of transliterated XLPT models before and after fine-tuning are more similar. 

Due to the less change of representation during fine-tuning, transliterated ZSDA has less forgetting of the pre-training knowledge, achieving better knowledge transfer across languages compared with the self-supervised ZSDA.

\subsection{Importance of accurate transliterations}
\label{sec:accurate}

The curriculum XLPT and the continuous pseudo-labeling in transliterated XLPT are proposed to improve the quality of pre-training labels, i.e., transliterations. Therefore, we use BT-CTC loss to examine how the above two techniques affect the quality of transliterations. The transliterations are generated on the source language data (CV Mandarin) using the target language (Cantonese) transliterated XLPT models. We select three different transliterated XLPT models: the proposed transliterated XLPT model, the one w/o curriculum XLPT, and the one w/o continuous pseudo-labeling (i.e., use the supervised seed model to generate transliterations). We also report the topline and baseline values of BT-CTC. The topline value is the CTC loss between the matched speech-transcription pairs of the source language, which is the case where the transliterations can accurately represent the content of the speech. And the baseline value is the CTC loss between mismatched speech-transcription pairs of the source language, which means the transliterations cannot represent the content of the speech.

As shown in \autoref{tab:BT-CTC}, we can observe that curriculum XLPT and continuous pseudo-labeling can effectively improve the quality of transliterations. And the fact that the BT-CTC values computed with transliterated XLPT models are between the baseline and topline values reflects that transliterations from the transliterated XLPT models can represent the content of the speech to some extent.

\begin{table}[htbp]
  \small
  \centering
  \caption{Measure the quality of transliterations with BT-CTC.}
    \begin{tabular}{lc}
    \toprule
    Method & BT-CTC \\
    \midrule
    Baseline & 243.0 \\
    \midrule
    Proposed   & 93.0 \\
    - w/o curriculum & 100.4 \\
    - w/o continuous pseudo-labeling & 100.9 \\
    \midrule
    Topline & 31.2 \\
    \bottomrule
    \end{tabular}%
  \label{tab:BT-CTC}%
\end{table}%

Then, we evaluate the performance of transliterated ZSDA without these two techniques. As shown in \autoref{tab:ablation_accuracy}, both curriculum XLST and continuous pseudo-labeling are critical techniques for decent performance. Combing the results in \autoref{tab:BT-CTC}, we can infer that better transliterations can lead to better performance. 

\begin{table}[htbp]
  \small
  \centering
  \caption{The performance of transliterated ZSDA w/o techniques for improving quality of transliterations.}
    \begin{tabular}{lccc}
    \toprule
    \multirow{2}[3]{*}{Method} & \multicolumn{3}{c}{Cantonese} \\
\cmidrule{2-4}          & Dev   & Test  & Avg \\
    \midrule
    Proposed   & 30.5  & 32.6  & 31.6 \\
    - w/o curriculum & 31.6  & 34.4  & 33.0 \\
    - w/o continuous pseudo-labeling & 31.3 & 33.6 & 32.5 \\
    \bottomrule
    \end{tabular}%
  \label{tab:ablation_accuracy}%
\end{table}%

\subsection{Selection of source language data}
\label{sec:selection_source}

In transliterated ZSDA, we select the source language data with two principles: (1) the source language should be in the same language family as the target language, and (2) the source language data should be in the target domain. We analyze the importance of these two designs with Cantonese as the target language in \autoref{tab:ablation_data}.

\begin{table}[htbp]
\small
  \centering
  \caption{WER comparison with different unlabeled source language data.}
  \begin{threeparttable}
    \begin{tabular}{lccc}
    \toprule
    \multirow{2}[3]{*}{Type of the source language data} & \multicolumn{3}{c}{Cantonese} \\
\cmidrule{2-4}          & Dev   & Test  & Avg \\
    \midrule
    Target domain close language (CV Mandarin) & 30.5  & 32.6  & 31.6 \\
    Target domain distant language (CV English) & 31.3  & 33.7  & 32.5 \\
    Cross-domain close language (AISHELL-1) & 33.3  & 36.3  & 34.8 \\
    \bottomrule
    \end{tabular}%
        \begin{tablenotes} 
        \footnotesize   
        \item[$*$] The durations of the three unlabeled cross-lingual datasets are similar.
    \end{tablenotes}  
    \end{threeparttable}
  \label{tab:ablation_data}%
\end{table}%

Firstly, we replace the language of the source language data from the close language (Mandarin) to a distant language (English). The new source language data remains in the target domain (CV). We can observe that the performance has degenerated, proving the importance of the similarity between the source and target languages. 

Then, keeping the source language as the close language Mandarin, we change its domain from the target domain (CV) to another (AISHELL-1). The significant WER increment indicates that the domain knowledge shared across languages is essential for the effectiveness of transliterated ZSDA.

\subsection{Analysis of transliterated XLPT designs}
\label{sec:XLPT_designs}

We conduct the ablation study in \autoref{tab:ablation_T-XLPT} to show the importance of the two designs in transliterated XLPT: the separate final linear layers and the graphemic-level supervision.

\begin{table}[htbp]
\small
  \centering
  \caption{Ablation study of transliterated XLPT in terms of WER}
    \begin{tabular}{lccc}
    \toprule
    \multirow{2}[3]{*}{XLPT method} & \multicolumn{3}{c}{Cantonese} \\
\cmidrule{2-4}          & Dev   & Test  & Avg \\
\midrule
    Transliterated XLPT & 30.5  & 32.6  & 31.6 \\
    - w/o separated layers & 31.2  & 33.4  & 32.3 \\
    - graphemic $\rightarrow$ phonemic & 31.9  & 34.7  & 33.3 \\
    \bottomrule
    \end{tabular}%
  \label{tab:ablation_T-XLPT}%
\end{table}%

\begin{figure}[htbp]
    \centering
	\subfloat[Transcription] 
	{ \label{fig:transcription_distribution}
		\includegraphics[width=0.45\columnwidth]{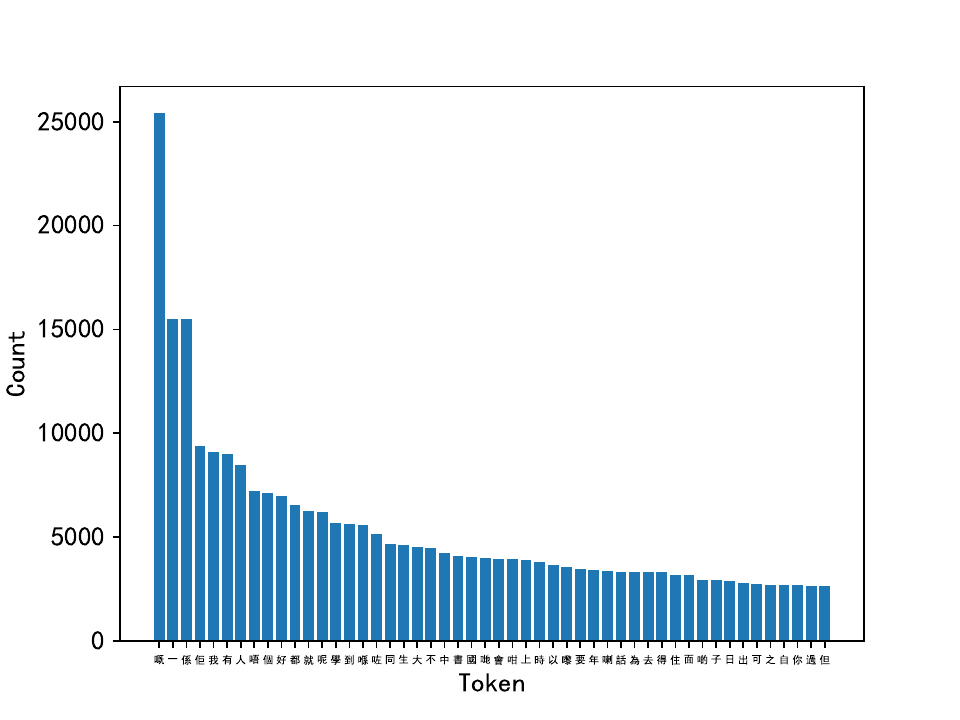}
	}\qquad
	\subfloat[Transliteration] 
	{ \label{fig:transliteration_distribution}
		\includegraphics[width=0.45\columnwidth]{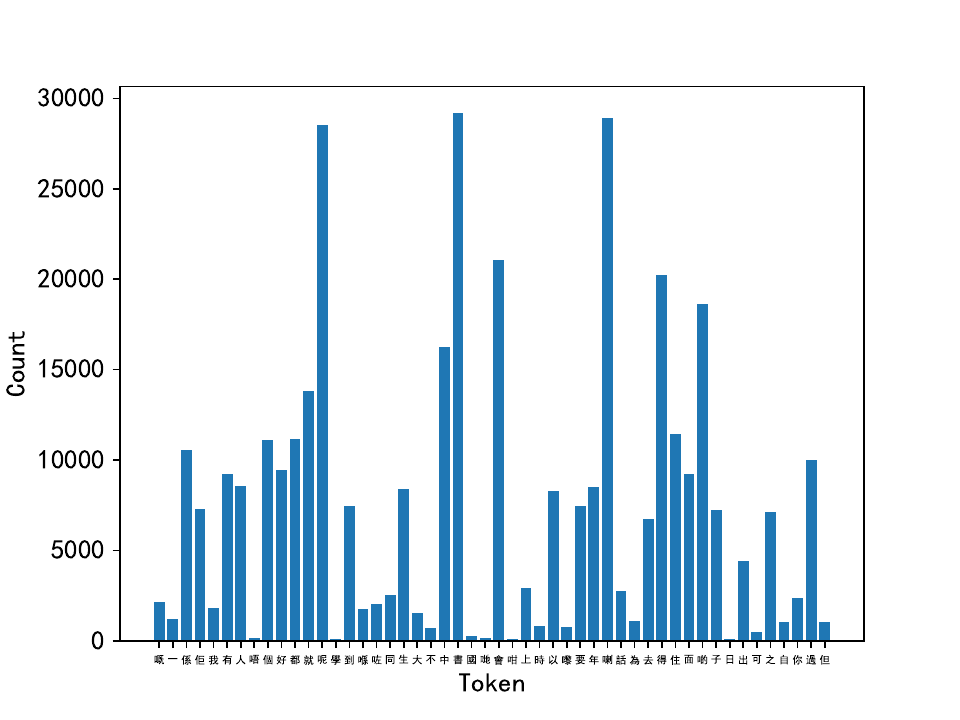}
	}
	\caption{Comparison of token distributions in transcriptions and transliterations.} 
	\label{fig:comparison_distribution}
\end{figure}

Transliterated XLPT without separate final linear layers would model the distributions of transcriptions and translations with the same parameters. The distribution gap (shown in \autoref{fig:comparison_distribution}) results in interference of optimization in the final linear layer, resulting in WER increment.

Then, when we switch the supervision type of transcriptions and transliterations from graphemic to phonemic, the performance is clearly degenerated. The reason is that the phonemic supervision would lead to inconsistency between the pre-training and fine-tuning labels.

\subsection{Analysis of curriculum XLPT}
\label{sec:curriculum}

\begin{table}[htbp]
\small
  \centering
  \caption{WER of transliterated ZSDA with variants of curriculum XLPT.}
    \begin{tabular}{lccc}
    \toprule
    \multirow{2}[3]{*}{Strategy} & \multicolumn{3}{c}{Cantonese} \\
\cmidrule{2-4}          & Dev   & Test  & Avg \\
    \midrule
    Baseline (wav2vec 2.0 on LibriSpeech) & 31.6  & 34.4  & 33.0 \\
    curriculum XLPT & 30.5  & 32.6  & 31.6 \\
    - w/o target language data & 31.2  & 33.5  & 32.4 \\
    - w/o source language data & 33.6  & 36.6  & 35.1 \\
    \bottomrule
    \end{tabular}%
  \label{tab:ablation_curriculum}%
\end{table}%

Finally, we analyze the effect of source/target language data in curriculum XLPT in \autoref{tab:ablation_curriculum}. The proposed curriculum XLPT method provides the best performance by initializing the model with self-supervised pre-training in both languages. Removing the target language data leads to slight performance degradation. Nonetheless, the improvement is still significant compared to the initialization with the LibriSpeech wav2vec 2.0 model, as the model is still pre-trained on the target domain source language distribution. However, when we remove the source language data, the performance is even worse than the initialization with the LibriSpeech wav2vec 2.0 model since it overfits the target language data by pre-training and fine-tuning with the same data.

\subsection{Summary of experimental results}

The key findings from experimental results are summarised as follows.
\begin{itemize}
    \item Self-supervised XLPT is an effective way for ZSDA (shown in \autoref{sec:comparison}), although it suffers critical representation change and knowledge forgetting during fine-tuning (shown in \autoref{sec:consistent}).
    \item Transliterated XLPT can minimize representation change and knowledge forgetting during fine-tuning by ensuring consistency between pre-training and fine-tuning labels (shown in \autoref{sec:consistent}), thus improving the performance of ZSDA (shown in \autoref{sec:comparison}).
    \item The effectiveness of transliterated XLPT depends on the accuracy of transliterations (\autoref{sec:accurate}), the similarity between source and target language (\autoref{sec:selection_source}), and some specific designs in transliterated XLPT (\autoref{sec:XLPT_designs} and \autoref{sec:curriculum}).
\end{itemize}

\section{Conclusion}
\label{sec:conclusion}
This work addresses the ZSDA problem by leveraging privileged information from cross-lingual target domain data. Specifically, we utilize the cross-lingual data via XLPT and target language fine-tuning. Since this practice suffers from the pre-trained knowledge forgetting issue during fine-tuning, we further offer the transliterated ZSDA method that innovatively utilizes transliteration-based pre-training labels during XLPT to minimize the representation change. Experimental results demonstrate that transliterated ZSDA realizes 9.2\% relative WER reduction on the target domain a wav2vec 2.0 baseline, without using target domain data in the target language. Furthermore, transliterated ZSDA consistently outperforms self-supervised ZSDA (with self-supervised pre-training labels) and has the same average performance with the supervised ZSDA (with ground-truth pre-training labels), demonstrating the advantage of transliteration-based pre-training labels.

\section{Acknowledgment}

This work is partially supported by the National Key R\&D Program of China (2022ZD0116103), the Youth Innovation Promotion Association, Chinese Academy of Sciences, and the Frontier Exploration Project Independently Deployed by Institute of Acoustics, Chinese Academy of Sciences under Grant QYTS202011.

\appendix

\section{Analysis of domain knowledge preservation}
\label{sec:knowlege_preservation_analysis}

We formally analyze the domain knowledge preservation of the pre-trained model during fine-tuning as follows.

Firstly, we formulate the training tasks in fine-tuning and pre-training. The fine-tuning task is to predict the transcription $\mathbf{y}$ given the speech $\mathbf{x}$, which can be decomposed to
\begin{equation}
P(\mathbf{y} \mid \mathbf{x})=\sum_\mathbf{z} P(\mathbf{y} \mid \mathbf{z}) P(\mathbf{z} \mid \mathbf{x})
\end{equation}
where $P(\mathbf{y} \mid \mathbf{x})$ is the entire ASR model, $P(\mathbf{z} \mid \mathbf{x})$ is the representation extractor, $P(\mathbf{y} \mid \mathbf{z})$ is the classifier, and $\mathbf{z}$ is the hidden representation.

Similarly, the pre-training task can be formulated as:
\begin{equation}
P(\mathbf{x}^{\prime} \mid \mathbf{x})=\sum_{\mathbf{z}^{\prime}} P(\mathbf{x}^{\prime} \mid \mathbf{z}^{\prime}) P(\mathbf{z}^{\prime} \mid \mathbf{x})
\end{equation}
where $\mathbf{x}^{\prime}$ is the self-supervised labels extracted from $\mathbf{x}$.

In the pre-training and fine-tuning paradigm, the representation extractor of the ASR model $P(\mathbf{z} \mid \mathbf{x})$ is initialized from the pre-trained representation extractor $P(\mathbf{z}^{\prime} \mid \mathbf{x})$. The classifier of the pre-trained model $P(\mathbf{x}^{\prime} \mid \mathbf{z}^{\prime})$ would be discarded after pre-training and the classifier of the ASR model is randomly initialized. 

The goal of pre-training is to provide a better initialization for the representation extractor. However, we can observe that the produced representations are mismatched between two representation extractors, i.e. $\mathbf{z}$ in the ASR model and $\mathbf{z}^{\prime}$ in the pre-trained model, which stems from the difference of the pre-training label $\mathbf{x}^{\prime}$ and the fine-tuning label $\mathbf{y}$. In self-supervised pre-training, the pre-training label comprises various low-level features of the speech. Consequently, the representation $\mathbf{z}^{\prime}$ should also retain such low-level features. In contrast, the fine-tuning label is the transcription thus the representation $\mathbf{z}$ is a high-level representation that only concentrate on the textual information. Such mismatch results in significant change of representations and parameters during fine-tuning. In the next, we illustrate why such change can lead to forgetting of the shared domain knowledge among languages and lead to insufficient domain knowledge transfer.

For illustration purpose, we focus on the last two layers of a cross-entropy (CE) based ASR model, i.e., the last layer of the representation extractor $\mathbf{W}_e \in \mathbb{R}^{E \times N}$ and the linear classification layer $\mathbf{W}_c \in \mathbb{R}^{C \times E}$. We omit the bias for simplicity. $\mathbf{o} \in \mathbb{R}^{N}$ is the representation before $\mathbf{W}_e$, $\mathbf{z} = \mathbf{W}_e \mathbf{o} \in \mathbb{R}^{E}$ is the extracted representation, $\mathbf{v} = \mathbf{W}_c \mathbf{z} \in \mathbb{R}^{C}$ is the predicted vector, and $\mathbf{y} \in \mathbb{R}^{C}$ is the one-hot label vector. 

For each frame, we can formulate the CE loss function as: 

\begin{equation}
L_{\mathrm{CE}}(\mathbf{o}, \mathbf{y} ; \mathbf{W}_e, \mathbf{W}_c)=- \mathbf{y} \cdot \log \operatorname{softmax} (\mathbf{W}_c \mathbf{W}_e \mathbf{o})
\end{equation}

Then, the gradient w.r.t. $\mathbf{W}_c$ can be computed as:
\begin{equation}
\label{equ:w_c}
\frac{\partial L_{\mathrm{CE}}(\mathbf{o}, \mathbf{y} ; \mathbf{W}_e, \mathbf{W}_c)}{\partial \mathbf{W}_c}=\frac{\partial L_{\mathrm{CE}}(\mathbf{o}, \mathbf{y} ; \mathbf{W}_e, \mathbf{W}_c)}{\partial \mathbf{v}} \frac{\partial \mathbf{v}}{\partial \mathbf{W}_c}
\end{equation}
where
\begin{equation}
\begin{aligned}
\frac{\partial L_{\mathrm{CE}}(\mathbf{o}, \mathbf{y} ; \mathbf{W}_e, \mathbf{W}_c)}{\partial \mathbf{v}} 
&= - \frac{\partial \mathbf{y} \cdot \log \operatorname{softmax}\left(\mathbf{v}\right)}{\partial \mathbf{v}} \\
& =-\frac{\partial \sum_{i=1}^C y_i \log \operatorname{softmax}_i\left(\mathbf{v}\right)}{\partial \mathbf{v}} \\
& =\frac{\partial \sum_{i=1}^C y_i \log \sum_{j=1}^C e^{v_j}}{\partial \mathbf{v}}-\frac{\partial \sum_{i=1}^C y_i \log e^{v_i}}{\partial \mathbf{v}} \\
& =\frac{\partial \log \sum_{j=1}^C e^{v_j}}{\partial \mathbf{v}}-\frac{\partial \sum_{i=1}^C y_i v_i}{\partial \mathbf{v}} \\
& =\left[\begin{array}{c}
\frac{e_1^{v_1}}{\sum_{j=1}^C e^{v_j}} \\
\vdots \\
\frac{e^{v_i}}{\sum_{j=1}^C e^{v_j}} \\
\vdots \\
\frac{e^{v_C}}{\sum_{j=1}^C e^{v_j}}
\end{array}\right]-\left[\begin{array}{c}
y_1 \\
\vdots \\
y_i \\
\vdots \\
y_C
\end{array}\right] \\
& =\operatorname{softmax}(\mathbf{v}) - \mathbf{y}
\end{aligned}
\end{equation}
and

\begin{equation}
\frac{\partial \mathbf{v}}{\partial \mathbf{W}_c}= \mathbf{z}^\mathrm{T}
\end{equation}

Therefore, \autoref{equ:w_c} can be transformed into:
\begin{equation}
\frac{\partial L_{\mathrm{CE}}(\mathbf{o}, \mathbf{y} ; \mathbf{W}_e, \mathbf{W}_c)}{\partial \mathbf{W}_c}= (\operatorname{softmax}(\mathbf{v}) - \mathbf{y}) \mathbf{z}^\mathrm{T} = (\operatorname{softmax}(\mathbf{W}_c \mathbf{z}) - \mathbf{y}) \mathbf{z}^\mathrm{T}
\end{equation}

Similarly, we can also get
\begin{equation}
\frac{\partial L_{\mathrm{CE}}(\mathbf{o}, \mathbf{y} ; \mathbf{W}_e, \mathbf{W}_c)}{\partial \mathbf{W}_e} 
= \mathbf{W}_c^\mathrm{T} (\operatorname{softmax}(\mathbf{W}_c \mathbf{W}_e \mathbf{o} - \mathbf{y})  \mathbf{o}^\mathrm{T}
\end{equation}

For an extreme case where the representation of an out-of-domain speech sample is in the direction orthogonal to the representation subspace of the fine-tuning data, say $\mathbf{z}^{\prime} \in S_{\mathbf{z}}^{\perp}$ and $\mathbf{o}^{\prime} \in S_{\mathbf{o}}^{\perp}$. The gradients computed on the fine-tuning data do not modify the representation and the prediction of this out-of-domain speech sample, i.e., $\frac{\partial L_{\mathrm{CE}}}{\partial \mathbf{W}_e}\mathbf{o}^{\prime} = 0$ or $\frac{\partial L_{\mathrm{CE}}}{\partial \mathbf{W}_c}\mathbf{z}^{\prime} = 0$. Consequently, the optimization on the fine-tuning data can not guarantee the good performance on the out-of-domain data, even if the pre-trained representation extractor is perfect~\citep{kumarfine}. In our setting, the fine-tuning data only consists of source domain. Therefore, if the representation change during fine-tuning is severe, the target performance is at risk even the pre-trained model can generalize well on the target domain, which means the shared domain knowledge that is learnt during pre-training is distorted during fine-tuning.


\bibliographystyle{model5-names}\biboptions{authoryear}
\bibliography{main}

\end{document}